\documentclass[aip,apl,reprint,superscriptaddress,showkeys,nofootinbib,floatfix]{revtex4-1}

\usepackage[english]{babel}			
\usepackage[utf8]{inputenc}		
\usepackage{microtype}

\usepackage{color} 						
\definecolor{red}{rgb}{1,0,0}			
\definecolor{blue}{rgb}{0,0,1}			

\usepackage{natmove} 

\usepackage{amsmath,amssymb,amsfonts}	
\usepackage{array} 

\usepackage{graphicx}			
\usepackage{booktabs} 
\usepackage{subfigure} 
\usepackage{paralist} 

\usepackage{verbatim} 
\usepackage{pdfpages} 
\includepdfset{pagecommand=\thispagestyle{plain}} 

\usepackage[pdftex]{hyperref}  
\definecolor{dullmagenta}{rgb}{0.4,0,0.4}   
\definecolor{darkblue}{rgb}{0,0,0.4}
\definecolor{medblue}{rgb}{0,0,0.8}
\hypersetup{
	colorlinks=true, 		
	breaklinks=true,		
	linkcolor=medblue,
	citecolor=medblue,
	filecolor=dullmagenta,
	urlcolor=blue
} 

\renewcommand{\url}[1]{\href{#1}{[URL]}} 

\usepackage{soul}
\definecolor{hlyellow}{rgb}{0.9,0.9,0}
\sethlcolor{hlyellow}

\usepackage{upgreek}				

\newcommand{\nm}{\text{ nm}}	

\newcommand{\eV}{\text{ eV}}      

\newcommand{\V}{\text{ V}}			

\newcommand{\pA}{\text{ pA}}      	

\newcommand{\uApum}{\text{ } \upmu \text{A/} \upmu \text{m}} 	
\newcommand{\pApum}{\text{ pA/} \upmu \text{m}} 	

\newcommand{\degC}{\text{ }^{\circ} \text{C}} 
\newcommand{\degK}{\text{ K}}	 

\newcommand{\mVpdec}{\text{ mV/decade}}	

\newcommand{\etal}{\textit{et al.}}

\newcommand{\figref}[1]{Fig.\ \ref{#1}} 
\newcommand{\figletter}[1]{(\mbox{#1})}

\newcommand{\CF}{C$_4$F$_8$}
\newcommand{\SF}{SF$_6$}
\newcommand{\aSi}{\mbox{a-Si}}
\newcommand{\polySi}{\mbox{poly-Si}}
\newcommand{\SiO}{SiO$_2$}

\newcommand{\Vds}{V_\text{DS}}
\newcommand{\Vg}{V_\text{G}}

\newcommand{\Ids}{I_\text{DS}}

\newcommand{\Ion}{I_\text{on}}
\newcommand{\Ioff}{I_\text{off}}

\newcommand{\Eadd}{E_\text{add}}
\newcommand{\Ec}{E_\text{C}}
\newcommand{\Ek}{E_\text{K}}

\begin{document}

\newcommand{\mytitle}
{A Silicon Nanocrystal Tunnel Field Effect Transistor}
\title{\mytitle}

\author{Patrick \surname{Harvey-Collard}}
\email[Corresponding author: ]{P.Collard@USherbrooke.ca; {Tel.: +1 819-821-8000 x65402}; {Fax: +1 819-821-8046}}
\affiliation{Département de physique, Université de Sherbrooke, 2500 boul.\ de l'Université, Sherbrooke, QC, J1K 2R1, Canada}
\affiliation{Institut Interdisciplinaire d'Innovation Technologique (3IT), Université de Sherbrooke, 3000 boul.\ de l'Université, Sherbrooke, QC, J1K 0A5, Canada}
\affiliation{Laboratoire Nanotechnologies Nanosystèmes (LN2)--CNRS UMI-3463, Université de Sherbrooke, 3000 boul.\ de l'Université, Sherbrooke, QC, J1K 0A5,  Canada}

\author{Dominique \surname{Drouin}}
\affiliation{Institut Interdisciplinaire d'Innovation Technologique (3IT), Université de Sherbrooke, 3000 boul.\ de l'Université, Sherbrooke, QC, J1K 0A5, Canada}
\affiliation{Laboratoire Nanotechnologies Nanosystèmes (LN2)--CNRS UMI-3463, Université de Sherbrooke, 3000 boul.\ de l'Université, Sherbrooke, QC, J1K 0A5,  Canada}

\author{Michel \surname{Pioro-Ladrière}}
\affiliation{Département de physique, Université de Sherbrooke, 2500 boul.\ de l'Université, Sherbrooke, QC, J1K 2R1, Canada}
\affiliation{CIFAR Program in Quantum Information Science, Canadian Institute for Advanced Research (CIFAR), Toronto, ON, M5G 1Z8, Canada}

\date{May 14$^\text{th}$, 2014}

\begin{abstract}
In this work, we demonstrate a silicon nanocrystal Field Effect Transistor (ncFET). Its operation is similar to that of a Tunnelling Field Effect Transistor (TFET) with two barriers in series. The tunnelling barriers are fabricated in very thin silicon dioxyde and the channel in intrinsic polycrystalline silicon. The absence of doping eliminates the problem of achieving sharp doping profiles at the junctions, which has proven a challenge for  large-scale integration and in principle allows scaling down the atomic level.
The demonstrated ncFET features a $10^4$ on/off current ratio at room temperature, a low $30\pApum$ leakage current at a $0.5\V$ bias, an on-state current on a par with typical all-Si TFETs and bipolar operation with high symmetry. Quantum dot transport spectroscopy is used to assess the band structure and energy levels of the silicon island.
\end{abstract}



\maketitle


The Complementary Metal--Oxide--Semiconductor (CMOS) technology is at the heart of modern electronic devices \cite{sze2007}. The performance of its fundamental component, the Metal--Oxide--Semiconductor Field Effect Transistor (MOSFET), can be evaluated in terms of on/off output current ratio, off-state leakage current, and subthreshold slope \cite{ferain2011}. However, certain performance characteristics of the MOSFET are intrinsically limited by the physics involved in the device \cite{sze2007,ferain2011,ionescu2011}. For example, the thermal activation of carriers imposes a limit on the subthreshold slope, which cannot be inferior to $60\mVpdec$ \cite{ferain2011}. A promising approach is to take advantage of the exponential dependance of tunnelling current on the width of a barrier to block thermal electrons while enabling a fast transistor turn-on with gate voltage, as it is the case in tunnelling Field Effect Transistors (TFETs) \cite{ionescu2011} or Schottky barrier Field Effect Transistors \cite{larson2006}. In typical TFETs, a single tunnelling barrier is formed by doping the source, channel and drain to form a p--i--n-like junction \cite{ionescu2011}. However, tight control of the doping profile is challenging because of dopant diffusion problems that prevent forming atomically sharp junctions \cite{pierre2010}, therefore limiting the built-in electric field and the on-state current \cite{seabaugh2010}.

\begin{figure}[tbp]
   \centering
   \includegraphics{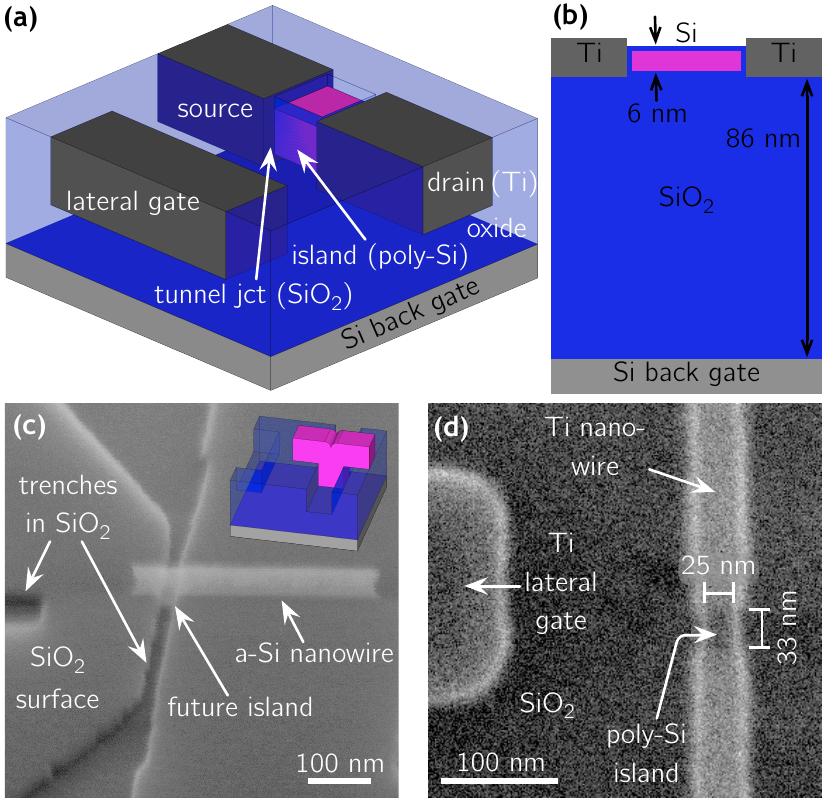} 
   \caption{\figletter{a} Schematic view of the ncFET (not to scale). \figletter{b} Schematic cross-section along the source/drain axis, showing transistor materials stack and actual dimensions. \figletter{c} Lateral-view Scanning Electron Microscope (SEM) image of a transistor at an intermediate fabrication step. It shows the source/drain nanotrench, the lateral gate trench and the etched \aSi{} nanowire overlapping the source/drain trench to form the future transistor island. \figletter{d} Top-view SEM image of a finished device. The final CMP step isolates the island after removing all material not embedded in a trench.}
   \label{fig:schematic}
\end{figure}

In this work, we fabricated and characterized a transistor made out of an intrinsic silicon nanocrystal contacted by two metallic source/drain electrodes, which we call a nanocrystal Field Effect Transistor (ncFET, \figref{fig:schematic}a). 
The switching mechanism relies on the gate-induced modulation of a tunnel barrier in a process very similar to a Schottky barrier FET \cite{larson2006}. This can, in principle, allow sub-60-mV/decade subthreshold slope and extremely low leakage current. Unlike many previous demonstrations of similar devices with self-assembled quantum dots \cite{klein1997,katsaros2010,lachance-quirion2014}, ours is entirely nanofabricated with industry-compatible techniques. The ncFET also differs from Schottky barrier FETs with silicide source/drain \cite{larson2006,matheu2012} by three facts. First, the leads are not made of a silicide but of an elemental metal. Second, the channel is very small in all spatial dimensions and separated from the leads by a thin insulator, which makes it effectively an island. Third, its very flexible fabrication process uses polycrystalline silicon (\polySi{}) and differs wildly from planar \cite{sze2007} or silicon-on-insulator \cite{ferain2011,roche2012} MOSFETs. Doping is intentionally avoided to enable extreme scaling of the device dimensions and avoid the dopant diffusion and non-uniformity problems mentioned earlier. 

The transistors were fabricated using a nanodamascene process based on the one of Dubuc \etal{} \cite{dubuc2008}. An oxidized silicon substrate is patterned with electron beam lithography and a CF$_4$/H$_2$/He Inductively Coupled Plasma (ICP) process \cite{guilmain2011} to produce $25\nm$ wide and $20\nm$ deep source/drain nanotrench and lateral gate trench. A 40-nm-thick amorphous silicon (\aSi{}) film is deposited using Low Pressure Chemical Vapor Deposition (LPCVD) at $525\degC$. An \aSi{} nanowire is then patterned over and perpendicular to the nanotrench using a \CF{}/\SF{} ICP etch \cite{harvey-collard2013}. The resulting structure is shown in \figref{fig:schematic}c. A $800\degC$ Rapid Thermal Anneal (RTA) is used to form \polySi{} with a thin film grain size in the range of $150$ to $300\nm$ with the goal of forming an island with few or no grains. The tunnel junctions are prepared by etching the native oxide at the surface of the \polySi{} nanowire using diluted hydrofluoric acid (HF) and letting it re-oxidize for 2 hours in cleanroom air before Ti deposition, which yields an estimated \SiO{} thickness of $0.5\nm$. Titanium is deposited and then polished using a Chemical Mechanical Polishing (CMP) process \cite{guilmain2013}, which isolates the silicon island and yields the structure of \figref{fig:schematic}d. This key step enables the unique nanocrystal geometry and positioning. Very interestingly, the nanodamascene \polySi{} process opens the door to using different lead and channel materials, as well as integration over various types of substrates.

\begin{figure}[tbp]
   \centering
   \includegraphics{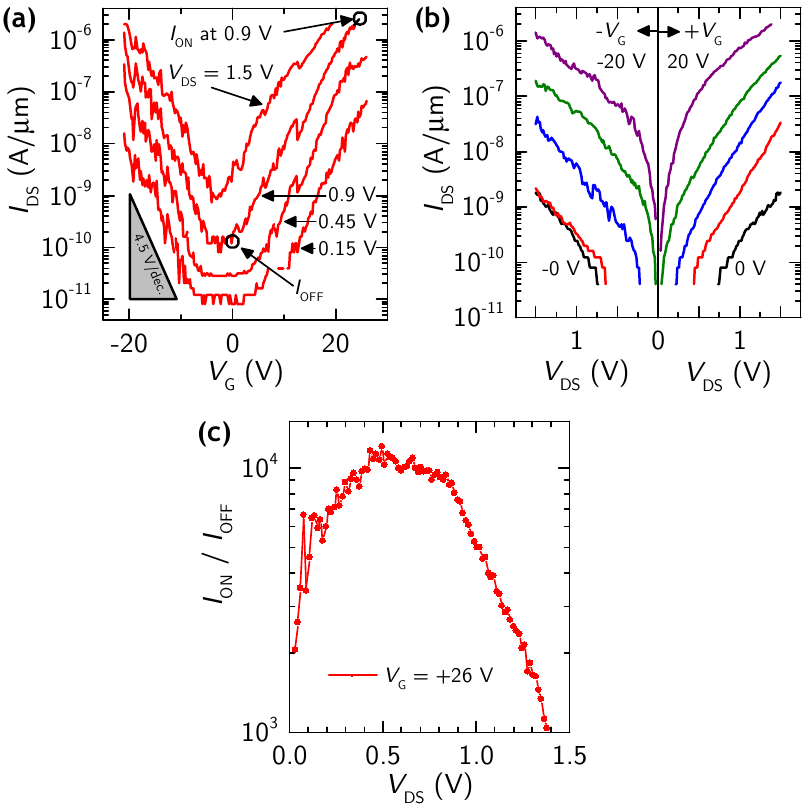} 
   \caption{\figletter{a} $\Ids-\Vg$ characteristics for $\Vds=0.15\V$, $0.45\V$, $0.9\V$ and $1.5\V$ showing near-perfect bipolar behavior. The subthreshold slope for this gate geometry is $4500 \mVpdec$. \figletter{b} $\Ids-\Vds$ characteristics for positive and negative values of $\Vg$ ($5\V$ intervals). \figletter{c} Demonstrated on/off current ratio as a function of $\Vds$. The $\Ion$ and $\Ioff$ values are taken from (a) at $\Vg=+26\V$ and $0\V$, respectively. The ncFET can therefore operate over a wide range of $\Vds$ values, making it capable of working at typical ($\sim1\V$) and low-power ($<0.5\V$) supply voltage.}
   \label{fig:results}
\end{figure}

The transistor characteristics of the ncFET are shown in \figref{fig:results}. 
Current through all terminals is monitored to validate that source--drain current goes through the island and that gate leakage is always below $0.5\pA$ and therefore negligible. The substrate is used as a back gate to supply voltage $\Vg$ relative to the drain with an oxide thickness of $86\nm$. The lateral gate being located $180\nm$ from the island in the featured device, its contribution is small compared with the back gate and hence is neglected in the following discussion. The channel width is $25\nm$. 

As shown by the data of \figref{fig:results}a, the transistor is a normally off device and works equally well for positive and negative gate biases. The current rises exponentially with gate bias, a clear sign of tunnelling. This rise shows no indication of saturation for all  accessible gate biases. Since the source is a metal, supply of electrons should not be a limiting factor. Moreover, the junction being in principle close to atomically sharp, it can build much bigger electric fields than doped junctions \cite{seabaugh2010}. Therefore, better gate electrostatics (e.g. lower equivalent oxide thickness and source--gate overlap) are expected to yield significantly better $\Ion$ \cite{loh2010}, while the demonstrated value of $2.5\uApum$ at $\Vds=0.9\V$ and $\Vg=25\V$ is already on a par with all-silicon TFETs that do not include CMOS technology boosters and source--gate overlap \cite{loh2010,ionescu2011}. The demonstrated $\Ion$ and $\Ioff$ are both roughly two orders of magnitude smaller than Schottky barrier FETs \cite{larson2006}. Taking into account that $\Ion$ could be further improved to $150\uApum$ (see discussion below), the ncFET then compares very favorably with both Schottky barrier FETs \cite{larson2006} and optimized all-silicon TFETs \cite{loh2010,ionescu2011}.

A plateau of leakage current is observed at low $\Vg$ and $\Vds$ (\figref{fig:results}a). Possible leakage mechanisms are thermionic emission over the barrier\cite{seabaugh2010}, Shockley-Read-Hall generation\cite{seabaugh2010} at the lead--island interface, and direct or trap-assisted tunnelling through the island \cite{seabaugh2010}. Preliminary results at low temperature on similar devices indicate that the leakage plateau is mostly independent of temperature. Therefore, the most likely leakage mechanism is direct and potentially trap-assisted tunnelling. This would also be consistent with the high current noise and will be discussed again later. The leakage current is $120\pApum$ at $\Vds=0.9\V$. This is 40 times smaller than low operating power MOSFETs and 830 times smaller than high performance MOSFETs \cite{ITRS2011-PIDS}. The leakage current (extracted from \figref{fig:results}a at $\Vg=0\V$) reduces to $30 \pApum$ at a lower $\Vds=0.45\V$ value, which is the projected low-operating-power supply voltage \cite{ITRS2011-PIDS}.

An important parameter is the $\Ion/\Ioff$ ratio. From the graph of \figref{fig:results}c, it is seen that $\Ion/\Ioff$ is maintained well above $10^3$ over a wide range of $\Vds$ values, from $0\V$ to $1.5\V$, and reaches $1.2\times10^4$. This makes the ncFET a very versatile device able to operate both at high and low supply voltage. While improving $\Ioff$ would require a longer channel or an improved processing, $\Ion$ can be easily enhanced. In fact, redesigning the interconnect circuit could allow to push the gate swing to $30\V$, improving $\Ion/\Ioff$ to an extrapolated value of $10^6$ by increasing $\Ion$ to $150\uApum$, enough to start competing directly with the one of MOSFETs \cite{ionescu2011,ITRS2011-PIDS}. Such a high $\Ion$ has already been achieved using a more favorable gate configuration \cite{tsui2004}. Moreover, adding a state-of-the-art high-k dielectric gate stack could reduce the equivalent oxide thickness from $86\nm$ to below $1\nm$. This would correspond to a scaling of the gate voltage from $30\V$ to $0.35 \V$ and of the subthreshold slope from $4500\mVpdec$ to below $52\mVpdec$, making it useful for low-power applications.

\begin{figure}[tbp]
   \centering
   \includegraphics{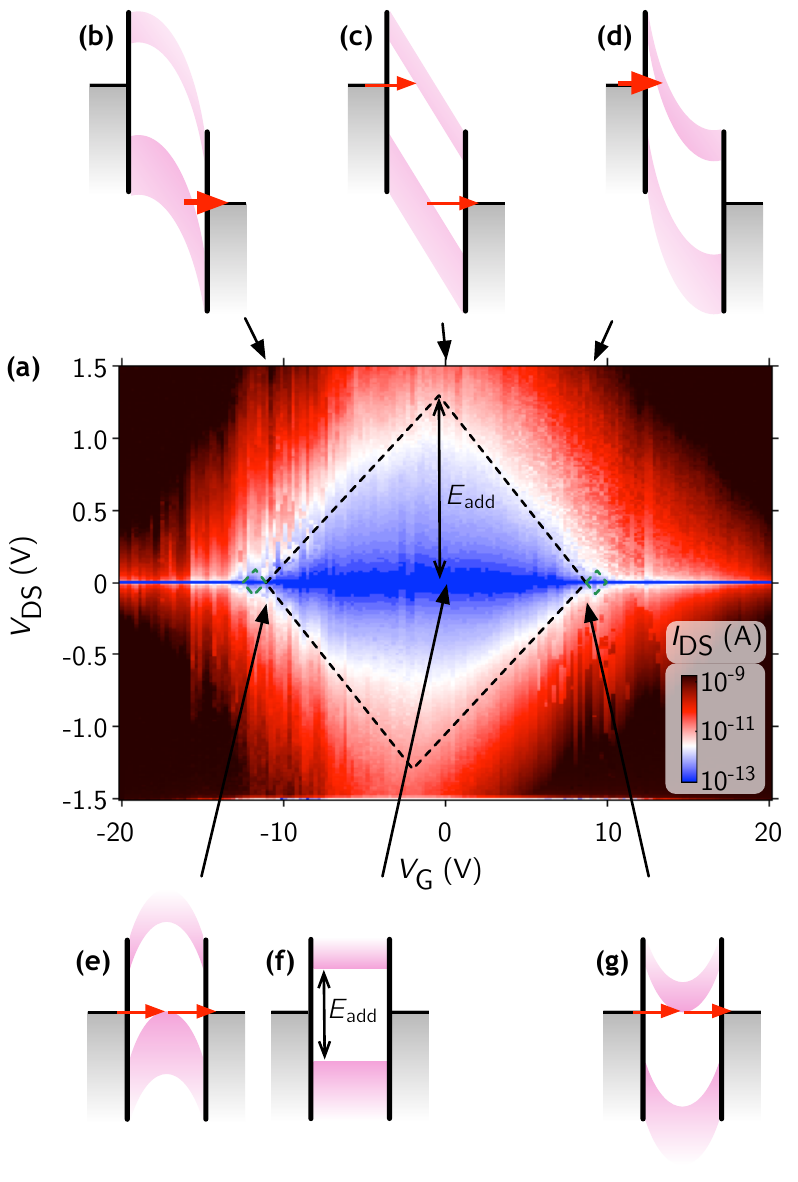} 
   \caption{\figletter{a} Transport spectroscopy at $300\degK$ showing features similar to Coulomb blockade (black dotted line). Here, the current is plotted in a logarithmic scale as a function of both $\Vds$ and $\Vg$. The large $\Eadd=1.3\eV$ is caused by the silicon bandgap, enlarged from its bulk value $1.12 \eV$ by strong quantum confinement in the vertical direction. Note that the current is not per unit gate width. \figletter{b--g} Band edge schematic for different values of gate voltage and source--drain bias. Diagrams are approximately to scale. The thick vertical black line represents the thin \SiO{} barrier. \figletter{f} Neutral band alignment, showing the slightly enlarged silicon bandgap. \figletter{e, g} At low $\Vds$, the gate attracts electrons (g) or holes (e) onto the island. The large quantum confinement energy causes additional hundred meV, ill-defined, room-temperature Coulomb diamonds (green dotted lines). \figletter{c} Leakage mechanism where electrons can tunnel through the island even in the absence of gate bias due to the small island length. \figletter{b, d} Gate bias changes the tunnel barrier width, allowing to maintain good $\Ion/\Ioff$ even at high $\Vds$.     }
   \label{fig:transportmech}
\end{figure}

We now investigate the similarity of the structure with quantum dots. This is motivated by the geometry of the device, which is expected to trap electrons inside the channel due to the thin oxide layer in the tunnel junction. In fact, quantum dot transport spectroscopy can provide us with insights on the band structure of the channel or island and help understand the transport mechanisms. In the quantum dot picture, the energy states of the electrons in the island are discrete due to the charging energy $\Ec$ and quantum confinement energy $\Ek$ \cite{likharev1999}. The energy required to add an electron onto the island is then $\Eadd = \Ec + \Ek$.
A transport spectroscopy measurement can allow to identify these energy levels \cite{hanson2007}. It consists of a measurement of $\Ids$ as a function of both $\Vds$ and $\Vg$. The diamond-shaped regions of blocked current, called Coulomb diamonds, contain the information on the energy level structure of the island. Figure \ref{fig:transportmech}a shows such a measurement, where a wide, diamond-shaped region of blocked current appears in the center of the plot. 
In our structure, the silicon gap appears as a large gap between the energy levels of the valence and conduction bands (\figref{fig:transportmech}f). If the temperature is higher than the small intraband levels, but much smaller than the gap, we then expect to see one large Coulomb diamond with an addition energy approximately equal to the silicon gap $1.12 \eV$ \cite{sze2007}. Our data does agree with this model. From \figref{fig:transportmech}a (black dashed line), we extract an addition energy of $1.3 \eV$, close but larger than the gap value. It is suspected that the difference could be accounted for by the very small dimension of the island in the vertical direction (around 5 to $7\nm$), where quantum confinement causes an increase in bandgap\cite{ma2003}. The high symmetry between positive and negative values of $\Vg$ is a strong indication that the band alignment of \figref{fig:transportmech}f is experimentally correct. This allows us to describe qualitatively the band structure for different configurations of $\Vds$ and $\Vg$. For example, the diagram of \figref{fig:transportmech}c illustrates the main leakage mechanism at high $\Vds$. When gate bias is applied (\figref{fig:transportmech}b or \ref{fig:transportmech}d), the bands are curved up or down, allowing to modulate the width of the tunnelling barrier. The electrostatic screening from the leads causes the gate to have more influence on the center of the channel than on its ends. This causes a band curvature like the one of \figref{fig:transportmech}e and \ref{fig:transportmech}g. Screening effects by the electrons in the channel are neglected, because they are expected to change only the quantitative behavior of the bands. In fact, in these bias configurations, the island is expected to form a potential well that would confine electrons or holes and produce Coulomb blockade at a lower temperature \cite{prati2012}. Indeed, the features observed at low $\Vds$ of \figref{fig:transportmech}a (green dashed lines) could be attributed to poorly confined and/or temperature-blurred Coulomb blockade. 

From the data of \figref{fig:results}a and \ref{fig:transportmech}a, it is seen that the noise level is very high and sometimes telegraphic in nature (not shown). Moreover, charge jumps are visible in the diamond of \figref{fig:transportmech}a. It is suspected that the number of charge defects is large, producing charge fluctuators \cite{zimmerman2007,zimmerman2008} or trap states that can affect the tunnel junctions. Because the tunnelling transport mechanism is extremely sensitive to distance, any small variation in the device's electrostatic environment is transduced in large current fluctuations. It is unclear whether the defects are located in the surrounding dielectric, the \SiO{} tunnel barrier or the island itself. Nevertheless, we do think that an anneal of the \SiO{} tunnel junction would be beneficial. It should be added that it is expected that the titanium of the leads will consume part of the tunnel junction over time. This process can lead to traps in the barrier and time-evolving behavior. In the future, replacing the \SiO{} and leads materials should be considered.
  
In summary, we have fabricated a tunnelling field effect transistor with metallic source and drain where the channel is made of a silicon nanocrystal. This transistor relies on intrinsic material for its operation, therefore eliminating the need for doping and allowing aggressive scaling down to the atomic level. Electronic transport is found to occur via tunnelling from the metal source to the channel conduction band at positive gate bias due to the strong electric field achieved in the absence of doping. A relatively good on/off current ratio, reaching $10^4$, is demonstrated at room temperature over a wide range of source--drain bias. We argue that this could be improved to $10^6$ with better gate electrostatics. Quantum dot transport spectroscopy was used to assess the band structure and energy levels of the silicon island. It shows that the ncFET could potentially be used as a single electron transistor and trap single electrons or holes. Future investigations will address the details of the transport mechanisms through low temperature transport measurements.

The authors would like to thank S. Ecoffey for help with chemical mechanical polishing and J.-P. Richard for preliminary work on a-Si/poly-Si. This work was supported by NSERC, FRQNT, NanoQuébec, RQMP and CIFAR.

%

\clearpage

\begin{thebibliography}{24}%
\makeatletter
\providecommand \@ifxundefined [1]{%
 \@ifx{#1\undefined}
}%
\providecommand \@ifnum [1]{%
 \ifnum #1\expandafter \@firstoftwo
 \else \expandafter \@secondoftwo
 \fi
}%
\providecommand \@ifx [1]{%
 \ifx #1\expandafter \@firstoftwo
 \else \expandafter \@secondoftwo
 \fi
}%
\providecommand \natexlab [1]{#1}%
\providecommand \enquote  [1]{``#1''}%
\providecommand \bibnamefont  [1]{#1}%
\providecommand \bibfnamefont [1]{#1}%
\providecommand \citenamefont [1]{#1}%
\providecommand \href@noop [0]{\@secondoftwo}%
\providecommand \href [0]{\begingroup \@sanitize@url \@href}%
\providecommand \@href[1]{\@@startlink{#1}\@@href}%
\providecommand \@@href[1]{\endgroup#1\@@endlink}%
\providecommand \@sanitize@url [0]{\catcode `\\12\catcode `\$12\catcode
  `\&12\catcode `\#12\catcode `\^12\catcode `\_12\catcode `\%12\relax}%
\providecommand \@@startlink[1]{}%
\providecommand \@@endlink[0]{}%
\providecommand \url  [0]{\begingroup\@sanitize@url \@url }%
\providecommand \@url [1]{\endgroup\@href {#1}{\urlprefix }}%
\providecommand \urlprefix  [0]{URL }%
\providecommand \Eprint [0]{\href }%
\providecommand \doibase [0]{http://dx.doi.org/}%
\providecommand \selectlanguage [0]{\@gobble}%
\providecommand \bibinfo  [0]{\@secondoftwo}%
\providecommand \bibfield  [0]{\@secondoftwo}%
\providecommand \translation [1]{[#1]}%
\providecommand \BibitemOpen [0]{}%
\providecommand \bibitemStop [0]{}%
\providecommand \bibitemNoStop [0]{.\EOS\space}%
\providecommand \EOS [0]{\spacefactor3000\relax}%
\providecommand \BibitemShut  [1]{\csname bibitem#1\endcsname}%
\let\auto@bib@innerbib\@empty
\bibitem [{\citenamefont {Sze}\ and\ \citenamefont {Ng}(2007)}]{sze2007}%
  \BibitemOpen
  \bibfield  {author} {\bibinfo {author} {\bibfnamefont {S.~M.}\ \bibnamefont
  {Sze}}\ and\ \bibinfo {author} {\bibfnamefont {K.~K.}\ \bibnamefont {Ng}},\
  }\href@noop {} {\emph {\bibinfo {title} {{Physics of Semiconductor
  Devices}}}},\ \bibinfo {edition} {3rd}\ ed.\ (\bibinfo  {publisher}
  {{Wiley-Interscience}},\ \bibinfo {year} {2007})\BibitemShut {NoStop}%
\bibitem [{\citenamefont {Ferain}, \citenamefont {Colinge},\ and\ \citenamefont
  {Colinge}(2011)}]{ferain2011}%
  \BibitemOpen
  \bibfield  {author} {\bibinfo {author} {\bibfnamefont {I.}~\bibnamefont
  {Ferain}}, \bibinfo {author} {\bibfnamefont {C.~A.}\ \bibnamefont {Colinge}},
  \ and\ \bibinfo {author} {\bibfnamefont {J.-P.}\ \bibnamefont {Colinge}},\
  }\href {http://dx.doi.org/10.1038/nature10676} {\bibfield  {journal}
  {\bibinfo  {journal} {Nature}\ }\textbf {\bibinfo {volume} {479}},\ \bibinfo
  {pages} {310} (\bibinfo {year} {2011})}\BibitemShut {NoStop}%
\bibitem [{\citenamefont {Ionescu}\ and\ \citenamefont
  {Riel}(2011)}]{ionescu2011}%
  \BibitemOpen
  \bibfield  {author} {\bibinfo {author} {\bibfnamefont {A.~M.}\ \bibnamefont
  {Ionescu}}\ and\ \bibinfo {author} {\bibfnamefont {H.}~\bibnamefont {Riel}},\
  }\href {http://dx.doi.org/10.1038/nature10679} {\bibfield  {journal}
  {\bibinfo  {journal} {Nature}\ }\textbf {\bibinfo {volume} {479}},\ \bibinfo
  {pages} {329} (\bibinfo {year} {2011})}\BibitemShut {NoStop}%
\bibitem [{\citenamefont {Larson}\ and\ \citenamefont
  {Snyder}(2006)}]{larson2006}%
  \BibitemOpen
  \bibfield  {author} {\bibinfo {author} {\bibfnamefont {J.}~\bibnamefont
  {Larson}}\ and\ \bibinfo {author} {\bibfnamefont {J.~P.}\ \bibnamefont
  {Snyder}},\ }\href {http://dx.doi.org/10.1109/TED.2006.871842} {\bibfield
  {journal} {\bibinfo  {journal} {IEEE Transactions on Electron Devices}\
  }\textbf {\bibinfo {volume} {53}},\ \bibinfo {pages} {1048} (\bibinfo {year}
  {2006})}\BibitemShut {NoStop}%
\bibitem [{\citenamefont {Pierre}\ \emph {et~al.}(2010)\citenamefont {Pierre},
  \citenamefont {Wacquez}, \citenamefont {Jehl}, \citenamefont {Sanquer},
  \citenamefont {Vinet},\ and\ \citenamefont {Cueto}}]{pierre2010}%
  \BibitemOpen
  \bibfield  {author} {\bibinfo {author} {\bibfnamefont {M.}~\bibnamefont
  {Pierre}}, \bibinfo {author} {\bibfnamefont {R.}~\bibnamefont {Wacquez}},
  \bibinfo {author} {\bibfnamefont {X.}~\bibnamefont {Jehl}}, \bibinfo {author}
  {\bibfnamefont {M.}~\bibnamefont {Sanquer}}, \bibinfo {author} {\bibfnamefont
  {M.}~\bibnamefont {Vinet}}, \ and\ \bibinfo {author} {\bibfnamefont
  {O.}~\bibnamefont {Cueto}},\ }\href
  {http://dx.doi.org/10.1038/nnano.2009.373} {\bibfield  {journal} {\bibinfo
  {journal} {Nat Nano}\ }\textbf {\bibinfo {volume} {5}},\ \bibinfo {pages}
  {133} (\bibinfo {year} {2010})}\BibitemShut {NoStop}%
\bibitem [{\citenamefont {Seabaugh}\ and\ \citenamefont
  {Zhang}(2010)}]{seabaugh2010}%
  \BibitemOpen
  \bibfield  {author} {\bibinfo {author} {\bibfnamefont {A.~C.}\ \bibnamefont
  {Seabaugh}}\ and\ \bibinfo {author} {\bibfnamefont {Q.}~\bibnamefont
  {Zhang}},\ }\href {http://dx.doi.org/10.1109/JPROC.2010.2070470} {\bibfield
  {journal} {\bibinfo  {journal} {Proceedings of the IEEE}\ }\textbf {\bibinfo
  {volume} {98}},\ \bibinfo {pages} {2095} (\bibinfo {year}
  {2010})}\BibitemShut {NoStop}%
\bibitem [{\citenamefont {Klein}\ \emph {et~al.}(1997)\citenamefont {Klein},
  \citenamefont {Roth}, \citenamefont {Lim}, \citenamefont {Alivisatos},\ and\
  \citenamefont {McEuen}}]{klein1997}%
  \BibitemOpen
  \bibfield  {author} {\bibinfo {author} {\bibfnamefont {D.~L.}\ \bibnamefont
  {Klein}}, \bibinfo {author} {\bibfnamefont {R.}~\bibnamefont {Roth}},
  \bibinfo {author} {\bibfnamefont {A.~K.~L.}\ \bibnamefont {Lim}}, \bibinfo
  {author} {\bibfnamefont {A.~P.}\ \bibnamefont {Alivisatos}}, \ and\ \bibinfo
  {author} {\bibfnamefont {P.~L.}\ \bibnamefont {McEuen}},\ }\href
  {http://dx.doi.org/10.1038/39535} {\bibfield  {journal} {\bibinfo  {journal}
  {Nature}\ }\textbf {\bibinfo {volume} {389}},\ \bibinfo {pages} {699}
  (\bibinfo {year} {1997})}\BibitemShut {NoStop}%
\bibitem [{\citenamefont {Katsaros}\ \emph {et~al.}(2010)\citenamefont
  {Katsaros}, \citenamefont {Spathis}, \citenamefont {Stoffel}, \citenamefont
  {Fournel}, \citenamefont {Mongillo}, \citenamefont {Bouchiat}, \citenamefont
  {Lefloch}, \citenamefont {Rastelli}, \citenamefont {Schmidt},\ and\
  \citenamefont {De~Franceschi}}]{katsaros2010}%
  \BibitemOpen
  \bibfield  {author} {\bibinfo {author} {\bibfnamefont {G.}~\bibnamefont
  {Katsaros}}, \bibinfo {author} {\bibfnamefont {P.}~\bibnamefont {Spathis}},
  \bibinfo {author} {\bibfnamefont {M.}~\bibnamefont {Stoffel}}, \bibinfo
  {author} {\bibfnamefont {F.}~\bibnamefont {Fournel}}, \bibinfo {author}
  {\bibfnamefont {M.}~\bibnamefont {Mongillo}}, \bibinfo {author}
  {\bibfnamefont {V.}~\bibnamefont {Bouchiat}}, \bibinfo {author}
  {\bibfnamefont {F.}~\bibnamefont {Lefloch}}, \bibinfo {author} {\bibfnamefont
  {A.}~\bibnamefont {Rastelli}}, \bibinfo {author} {\bibfnamefont {O.~G.}\
  \bibnamefont {Schmidt}}, \ and\ \bibinfo {author} {\bibfnamefont
  {S.}~\bibnamefont {De~Franceschi}},\ }\href
  {http://dx.doi.org/10.1038/nnano.2010.84} {\bibfield  {journal} {\bibinfo
  {journal} {Nat Nano}\ }\textbf {\bibinfo {volume} {5}},\ \bibinfo {pages}
  {458} (\bibinfo {year} {2010})}\BibitemShut {NoStop}%
\bibitem [{\citenamefont {Lachance-Quirion}\ \emph {et~al.}(2014)\citenamefont
  {Lachance-Quirion}, \citenamefont {Tremblay}, \citenamefont {Lamarre},
  \citenamefont {M{\'e}thot}, \citenamefont {Gingras}, \citenamefont
  {Camirand~Lemyre}, \citenamefont {Pioro-Ladri{\`e}re},\ and\ \citenamefont
  {Allen}}]{lachance-quirion2014}%
  \BibitemOpen
  \bibfield  {author} {\bibinfo {author} {\bibfnamefont {D.}~\bibnamefont
  {Lachance-Quirion}}, \bibinfo {author} {\bibfnamefont {S.}~\bibnamefont
  {Tremblay}}, \bibinfo {author} {\bibfnamefont {S.~A.}\ \bibnamefont
  {Lamarre}}, \bibinfo {author} {\bibfnamefont {V.}~\bibnamefont {M{\'e}thot}},
  \bibinfo {author} {\bibfnamefont {D.}~\bibnamefont {Gingras}}, \bibinfo
  {author} {\bibfnamefont {J.}~\bibnamefont {Camirand~Lemyre}}, \bibinfo
  {author} {\bibfnamefont {M.}~\bibnamefont {Pioro-Ladri{\`e}re}}, \ and\
  \bibinfo {author} {\bibfnamefont {C.~N.}\ \bibnamefont {Allen}},\ }\href
  {http://dx.doi.org/10.1021/nl404247e} {\bibfield  {journal} {\bibinfo
  {journal} {Nano Letters}\ }\textbf {\bibinfo {volume} {14}},\ \bibinfo
  {pages} {882} (\bibinfo {year} {2014})}\BibitemShut {NoStop}%
\bibitem [{\citenamefont {Matheu}(2012)}]{matheu2012}%
  \BibitemOpen
  \bibfield  {author} {\bibinfo {author} {\bibfnamefont {P.}~\bibnamefont
  {Matheu}},\ }\emph {\bibinfo {title} {Investigations of Tunneling for Field
  Effect Transistors}},\ \href
  {http://www.eecs.berkeley.edu/~tking/theses/pmatheu.pdf} {Ph.D. thesis},\
  \bibinfo  {school} {University of California at Berkeley} (\bibinfo {year}
  {2012})\BibitemShut {NoStop}%
\bibitem [{\citenamefont {Roche}\ \emph {et~al.}(2012)\citenamefont {Roche},
  \citenamefont {Voisin}, \citenamefont {Jehl}, \citenamefont {Wacquez},
  \citenamefont {Sanquer}, \citenamefont {Vinet}, \citenamefont {Deshpande},\
  and\ \citenamefont {Previtali}}]{roche2012}%
  \BibitemOpen
  \bibfield  {author} {\bibinfo {author} {\bibfnamefont {B.}~\bibnamefont
  {Roche}}, \bibinfo {author} {\bibfnamefont {B.}~\bibnamefont {Voisin}},
  \bibinfo {author} {\bibfnamefont {X.}~\bibnamefont {Jehl}}, \bibinfo {author}
  {\bibfnamefont {R.}~\bibnamefont {Wacquez}}, \bibinfo {author} {\bibfnamefont
  {M.}~\bibnamefont {Sanquer}}, \bibinfo {author} {\bibfnamefont
  {M.}~\bibnamefont {Vinet}}, \bibinfo {author} {\bibfnamefont
  {V.}~\bibnamefont {Deshpande}}, \ and\ \bibinfo {author} {\bibfnamefont
  {B.}~\bibnamefont {Previtali}},\ }\href {http://dx.doi.org/10.1063/1.3678042}
  {\bibfield  {journal} {\bibinfo  {journal} {Applied Physics Letters}\
  }\textbf {\bibinfo {volume} {100}},\ \bibinfo {eid} {032107} (\bibinfo {year}
  {2012})}\BibitemShut {NoStop}%
\bibitem [{\citenamefont {Dubuc}, \citenamefont {Beauvais},\ and\ \citenamefont
  {Drouin}(2008)}]{dubuc2008}%
  \BibitemOpen
  \bibfield  {author} {\bibinfo {author} {\bibfnamefont {C.}~\bibnamefont
  {Dubuc}}, \bibinfo {author} {\bibfnamefont {J.}~\bibnamefont {Beauvais}}, \
  and\ \bibinfo {author} {\bibfnamefont {D.}~\bibnamefont {Drouin}},\ }\href
  {http://dx.doi.org/10.1109/TNANO.2007.913430} {\bibfield  {journal} {\bibinfo
   {journal} {IEEE Transactions on Nanotechnology}\ }\textbf {\bibinfo {volume}
  {7}},\ \bibinfo {pages} {68} (\bibinfo {year} {2008})}\BibitemShut {NoStop}%
\bibitem [{\citenamefont {Guilmain}\ \emph {et~al.}(2011)\citenamefont
  {Guilmain}, \citenamefont {Jaouad}, \citenamefont {Ecoffey},\ and\
  \citenamefont {Drouin}}]{guilmain2011}%
  \BibitemOpen
  \bibfield  {author} {\bibinfo {author} {\bibfnamefont {M.}~\bibnamefont
  {Guilmain}}, \bibinfo {author} {\bibfnamefont {A.}~\bibnamefont {Jaouad}},
  \bibinfo {author} {\bibfnamefont {S.}~\bibnamefont {Ecoffey}}, \ and\
  \bibinfo {author} {\bibfnamefont {D.}~\bibnamefont {Drouin}},\ }\bibfield
  {booktitle} {\emph {\bibinfo {booktitle} {Proceedings of the 36th
  International Conference on Micro- and Nano-Engineering (MNE)}},\ }\href
  {http://www.sciencedirect.com/science/article/pii/S0167931711001407}
  {\bibfield  {journal} {\bibinfo  {journal} {Microelectronic Engineering}\
  }\textbf {\bibinfo {volume} {88}},\ \bibinfo {pages} {2505} (\bibinfo {year}
  {2011})}\BibitemShut {NoStop}%
\bibitem [{\citenamefont {Harvey-Collard}\ \emph {et~al.}(2013)\citenamefont
  {Harvey-Collard}, \citenamefont {Jaouad}, \citenamefont {Drouin},\ and\
  \citenamefont {Pioro-Ladri{\`e}re}}]{harvey-collard2013}%
  \BibitemOpen
  \bibfield  {author} {\bibinfo {author} {\bibfnamefont {P.}~\bibnamefont
  {Harvey-Collard}}, \bibinfo {author} {\bibfnamefont {A.}~\bibnamefont
  {Jaouad}}, \bibinfo {author} {\bibfnamefont {D.}~\bibnamefont {Drouin}}, \
  and\ \bibinfo {author} {\bibfnamefont {M.}~\bibnamefont
  {Pioro-Ladri{\`e}re}},\ }\href {http://dx.doi.org/10.1016/j.mee.2013.02.099}
  {\bibfield  {journal} {\bibinfo  {journal} {Microelectronic Engineering}\
  }\textbf {\bibinfo {volume} {110}},\ \bibinfo {pages} {408} (\bibinfo {year}
  {2013})}\BibitemShut {NoStop}%
\bibitem [{\citenamefont {Guilmain}\ \emph {et~al.}(2013)\citenamefont
  {Guilmain}, \citenamefont {Labbaye}, \citenamefont {Dellenbach},
  \citenamefont {Nauenheim}, \citenamefont {Drouin},\ and\ \citenamefont
  {Ecoffey}}]{guilmain2013}%
  \BibitemOpen
  \bibfield  {author} {\bibinfo {author} {\bibfnamefont {M.}~\bibnamefont
  {Guilmain}}, \bibinfo {author} {\bibfnamefont {T.}~\bibnamefont {Labbaye}},
  \bibinfo {author} {\bibfnamefont {F.}~\bibnamefont {Dellenbach}}, \bibinfo
  {author} {\bibfnamefont {C.}~\bibnamefont {Nauenheim}}, \bibinfo {author}
  {\bibfnamefont {D.}~\bibnamefont {Drouin}}, \ and\ \bibinfo {author}
  {\bibfnamefont {S.}~\bibnamefont {Ecoffey}},\ }\href
  {http://stacks.iop.org/0957-4484/24/i=24/a=245305} {\bibfield  {journal}
  {\bibinfo  {journal} {Nanotechnology}\ }\textbf {\bibinfo {volume} {24}},\
  \bibinfo {pages} {245305} (\bibinfo {year} {2013})}\BibitemShut {NoStop}%
\bibitem [{\citenamefont {Loh}\ \emph {et~al.}(2010)\citenamefont {Loh},
  \citenamefont {Jeon}, \citenamefont {Kang}, \citenamefont {Oh}, \citenamefont
  {Patel}, \citenamefont {Smith}, \citenamefont {Barnett}, \citenamefont
  {Park}, \citenamefont {Liu}, \citenamefont {Tseng}, \citenamefont {Majhi},
  \citenamefont {Jammy},\ and\ \citenamefont {Hu}}]{loh2010}%
  \BibitemOpen
  \bibfield  {author} {\bibinfo {author} {\bibfnamefont {W.-Y.}\ \bibnamefont
  {Loh}}, \bibinfo {author} {\bibfnamefont {K.}~\bibnamefont {Jeon}}, \bibinfo
  {author} {\bibfnamefont {C.-Y.}\ \bibnamefont {Kang}}, \bibinfo {author}
  {\bibfnamefont {J.}~\bibnamefont {Oh}}, \bibinfo {author} {\bibfnamefont
  {P.}~\bibnamefont {Patel}}, \bibinfo {author} {\bibfnamefont
  {C.}~\bibnamefont {Smith}}, \bibinfo {author} {\bibfnamefont
  {J.}~\bibnamefont {Barnett}}, \bibinfo {author} {\bibfnamefont
  {C.}~\bibnamefont {Park}}, \bibinfo {author} {\bibfnamefont {T.-J.~K.}\
  \bibnamefont {Liu}}, \bibinfo {author} {\bibfnamefont {H.-H.}\ \bibnamefont
  {Tseng}}, \bibinfo {author} {\bibfnamefont {P.}~\bibnamefont {Majhi}},
  \bibinfo {author} {\bibfnamefont {R.}~\bibnamefont {Jammy}}, \ and\ \bibinfo
  {author} {\bibfnamefont {C.}~\bibnamefont {Hu}},\ }in\ \href
  {http://dx.doi.org/10.1109/ESSDERC.2010.5618418} {\emph {\bibinfo {booktitle}
  {Proceedings of the European Solid-State Device Research Conference (ESSDERC
  2010)}}}\ (\bibinfo {year} {2010})\ pp.\ \bibinfo {pages}
  {162--165}\BibitemShut {NoStop}%
\bibitem [{ITR(2011)}]{ITRS2011-PIDS}%
  \BibitemOpen
  \enquote {\bibinfo {title} {International technology roadmap for
  semiconductors report 2011},}\ \ (\bibinfo  {publisher} {International
  Technology Roadmap for Semiconductors},\ \bibinfo {year} {2011})\ Chap.\
  \bibinfo {chapter} {Process Integration, Devices, and Structures
  ({PIDS})}\BibitemShut {NoStop}%
\bibitem [{\citenamefont {Tsui}\ and\ \citenamefont {Lin}(2004)}]{tsui2004}%
  \BibitemOpen
  \bibfield  {author} {\bibinfo {author} {\bibfnamefont {B.-Y.}\ \bibnamefont
  {Tsui}}\ and\ \bibinfo {author} {\bibfnamefont {C.-P.}\ \bibnamefont {Lin}},\
  }\href {http://dx.doi.org/10.1109/LED.2004.828980} {\bibfield  {journal}
  {\bibinfo  {journal} {IEEE Electron Device Letters}\ }\textbf {\bibinfo
  {volume} {25}},\ \bibinfo {pages} {430} (\bibinfo {year} {2004})}\BibitemShut
  {NoStop}%
\bibitem [{\citenamefont {Likharev}(1999)}]{likharev1999}%
  \BibitemOpen
  \bibfield  {author} {\bibinfo {author} {\bibfnamefont {K.}~\bibnamefont
  {Likharev}},\ }\href {http://dx.doi.org/10.1109/5.752518} {\bibfield
  {journal} {\bibinfo  {journal} {Proceedings of the IEEE}\ }\textbf {\bibinfo
  {volume} {87}},\ \bibinfo {pages} {606} (\bibinfo {year} {1999})}\BibitemShut
  {NoStop}%
\bibitem [{\citenamefont {Hanson}\ \emph {et~al.}(2007)\citenamefont {Hanson},
  \citenamefont {Kouwenhoven}, \citenamefont {Petta}, \citenamefont {Tarucha},\
  and\ \citenamefont {Vandersypen}}]{hanson2007}%
  \BibitemOpen
  \bibfield  {author} {\bibinfo {author} {\bibfnamefont {R.}~\bibnamefont
  {Hanson}}, \bibinfo {author} {\bibfnamefont {L.~P.}\ \bibnamefont
  {Kouwenhoven}}, \bibinfo {author} {\bibfnamefont {J.~R.}\ \bibnamefont
  {Petta}}, \bibinfo {author} {\bibfnamefont {S.}~\bibnamefont {Tarucha}}, \
  and\ \bibinfo {author} {\bibfnamefont {L.~M.~K.}\ \bibnamefont
  {Vandersypen}},\ }\href {http://dx.doi.org/10.1103/RevModPhys.79.1217}
  {\bibfield  {journal} {\bibinfo  {journal} {Rev. Mod. Phys.}\ }\textbf
  {\bibinfo {volume} {79}},\ \bibinfo {pages} {1217} (\bibinfo {year}
  {2007})}\BibitemShut {NoStop}%
\bibitem [{\citenamefont {Ma}\ \emph {et~al.}(2003)\citenamefont {Ma},
  \citenamefont {Lee}, \citenamefont {Au}, \citenamefont {Tong},\ and\
  \citenamefont {Lee}}]{ma2003}%
  \BibitemOpen
  \bibfield  {author} {\bibinfo {author} {\bibfnamefont {D.~D.~D.}\
  \bibnamefont {Ma}}, \bibinfo {author} {\bibfnamefont {C.~S.}\ \bibnamefont
  {Lee}}, \bibinfo {author} {\bibfnamefont {F.~C.~K.}\ \bibnamefont {Au}},
  \bibinfo {author} {\bibfnamefont {S.~Y.}\ \bibnamefont {Tong}}, \ and\
  \bibinfo {author} {\bibfnamefont {S.~T.}\ \bibnamefont {Lee}},\ }\href
  {http://dx.doi.org/10.1126/science.1080313} {\bibfield  {journal} {\bibinfo
  {journal} {Science}\ }\textbf {\bibinfo {volume} {299}},\ \bibinfo {pages}
  {1874} (\bibinfo {year} {2003})}\BibitemShut {NoStop}%
\bibitem [{\citenamefont {Prati}\ \emph {et~al.}(2012)\citenamefont {Prati},
  \citenamefont {Michielis}, \citenamefont {Belli}, \citenamefont {Cocco},
  \citenamefont {Fanciulli}, \citenamefont {Kotekar-Patil}, \citenamefont
  {Ruoff}, \citenamefont {Kern}, \citenamefont {Wharam}, \citenamefont
  {Verduijn}, \citenamefont {Tettamanzi}, \citenamefont {Rogge}, \citenamefont
  {Roche}, \citenamefont {Wacquez}, \citenamefont {Jehl}, \citenamefont
  {Vinet},\ and\ \citenamefont {Sanquer}}]{prati2012}%
  \BibitemOpen
  \bibfield  {author} {\bibinfo {author} {\bibfnamefont {E.}~\bibnamefont
  {Prati}}, \bibinfo {author} {\bibfnamefont {M.~D.}\ \bibnamefont
  {Michielis}}, \bibinfo {author} {\bibfnamefont {M.}~\bibnamefont {Belli}},
  \bibinfo {author} {\bibfnamefont {S.}~\bibnamefont {Cocco}}, \bibinfo
  {author} {\bibfnamefont {M.}~\bibnamefont {Fanciulli}}, \bibinfo {author}
  {\bibfnamefont {D.}~\bibnamefont {Kotekar-Patil}}, \bibinfo {author}
  {\bibfnamefont {M.}~\bibnamefont {Ruoff}}, \bibinfo {author} {\bibfnamefont
  {D.~P.}\ \bibnamefont {Kern}}, \bibinfo {author} {\bibfnamefont {D.~A.}\
  \bibnamefont {Wharam}}, \bibinfo {author} {\bibfnamefont {J.}~\bibnamefont
  {Verduijn}}, \bibinfo {author} {\bibfnamefont {G.~C.}\ \bibnamefont
  {Tettamanzi}}, \bibinfo {author} {\bibfnamefont {S.}~\bibnamefont {Rogge}},
  \bibinfo {author} {\bibfnamefont {B.}~\bibnamefont {Roche}}, \bibinfo
  {author} {\bibfnamefont {R.}~\bibnamefont {Wacquez}}, \bibinfo {author}
  {\bibfnamefont {X.}~\bibnamefont {Jehl}}, \bibinfo {author} {\bibfnamefont
  {M.}~\bibnamefont {Vinet}}, \ and\ \bibinfo {author} {\bibfnamefont
  {M.}~\bibnamefont {Sanquer}},\ }\href
  {http://stacks.iop.org/0957-4484/23/i=21/a=215204} {\bibfield  {journal}
  {\bibinfo  {journal} {Nanotechnology}\ }\textbf {\bibinfo {volume} {23}},\
  \bibinfo {pages} {215204} (\bibinfo {year} {2012})}\BibitemShut {NoStop}%
\bibitem [{\citenamefont {Zimmerman}\ \emph {et~al.}(2007)\citenamefont
  {Zimmerman}, \citenamefont {Simonds}, \citenamefont {Fujiwara}, \citenamefont
  {Ono}, \citenamefont {Takahashi},\ and\ \citenamefont
  {Inokawa}}]{zimmerman2007}%
  \BibitemOpen
  \bibfield  {author} {\bibinfo {author} {\bibfnamefont {N.~M.}\ \bibnamefont
  {Zimmerman}}, \bibinfo {author} {\bibfnamefont {B.~J.}\ \bibnamefont
  {Simonds}}, \bibinfo {author} {\bibfnamefont {A.}~\bibnamefont {Fujiwara}},
  \bibinfo {author} {\bibfnamefont {Y.}~\bibnamefont {Ono}}, \bibinfo {author}
  {\bibfnamefont {Y.}~\bibnamefont {Takahashi}}, \ and\ \bibinfo {author}
  {\bibfnamefont {H.}~\bibnamefont {Inokawa}},\ }\href
  {http://dx.doi.org/10.1063/1.2431778} {\bibfield  {journal} {\bibinfo
  {journal} {Applied Physics Letters}\ }\textbf {\bibinfo {volume} {90}},\
  \bibinfo {pages} {033507} (\bibinfo {year} {2007})}\BibitemShut {NoStop}%
\bibitem [{\citenamefont {Zimmerman}\ \emph {et~al.}(2008)\citenamefont
  {Zimmerman}, \citenamefont {Huber}, \citenamefont {Simonds}, \citenamefont
  {Hourdakis}, \citenamefont {Fujiwara}, \citenamefont {Ono}, \citenamefont
  {Takahashi}, \citenamefont {Inokawa}, \citenamefont {Furlan},\ and\
  \citenamefont {Keller}}]{zimmerman2008}%
  \BibitemOpen
  \bibfield  {author} {\bibinfo {author} {\bibfnamefont {N.~M.}\ \bibnamefont
  {Zimmerman}}, \bibinfo {author} {\bibfnamefont {W.~H.}\ \bibnamefont
  {Huber}}, \bibinfo {author} {\bibfnamefont {B.}~\bibnamefont {Simonds}},
  \bibinfo {author} {\bibfnamefont {E.}~\bibnamefont {Hourdakis}}, \bibinfo
  {author} {\bibfnamefont {A.}~\bibnamefont {Fujiwara}}, \bibinfo {author}
  {\bibfnamefont {Y.}~\bibnamefont {Ono}}, \bibinfo {author} {\bibfnamefont
  {Y.}~\bibnamefont {Takahashi}}, \bibinfo {author} {\bibfnamefont
  {H.}~\bibnamefont {Inokawa}}, \bibinfo {author} {\bibfnamefont
  {M.}~\bibnamefont {Furlan}}, \ and\ \bibinfo {author} {\bibfnamefont {M.~W.}\
  \bibnamefont {Keller}},\ }\href {http://dx.doi.org/10.1063/1.2949700}
  {\bibfield  {journal} {\bibinfo  {journal} {Journal of Applied Physics}\
  }\textbf {\bibinfo {volume} {104}},\ \bibinfo {eid} {033710} (\bibinfo {year}
  {2008})}\BibitemShut {NoStop}%
\end{thebibliography}
\end{document}